\documentclass[a4paper,10pt]{article}

\newcommand{\lyxaddress}[1]{
\par {\raggedright #1
\vspace{1.4em}
\noindent\par}
}
\usepackage{epsfig}
\usepackage[authoryear]{natbib}

\begin{document}
\title{The effect of neural adaptation of population coding accuracy}
\author{J.M. Cortes$^{1,5}$, D. Marinazzo$^{2}$, P. Series$^{1}$, M.W. Oram$^{4}$,T.J.
Sejnowski$^{3}$ \\ and M.C.W. van Rossum$^{1,6}$}

\maketitle

\lyxaddress{
1. Institute for Adaptive and Neural Computation. 
 School of Informatics. University of Edinburgh, UK \\
 2. Laboratory of Neurophysics and Physiology. 
 CNRS-Universite Rene Descartes, France\\
 3. Howard Hughes Medical Institute. 
 The Salk Institute, La Jolla CA, USA \\
4. School of Psychology, University of St Andrews, UK \\
5. Present Address: Departamento de Ciencias de la Computacion 
			 e Inteligencia Artificial. 
Universidad de Granada, Spain  \\
6. Correspondence to: Mark van Rossum. email: mvanross@inf.ed.ac.uk \\
}

\newpage

\begin{abstract}
Most neurons in the primary visual cortex initially respond vigorously when a
preferred stimulus is presented, but adapt as stimulation continues.
The functional consequences of adaptation are unclear. Typically a reduction of
firing rate would reduce single neuron accuracy as less spikes are available
for decoding, but it has been suggested that on the
population level, adaptation increases coding accuracy. This question
requires
careful analysis as adaptation not only changes the firing rates
of neurons, but also the neural variability and correlations between
neurons, which affect coding accuracy as well. We calculate the coding
accuracy using a computational model that
implements two forms of adaptation: spike frequency adaptation and synaptic
adaptation in the form of short-term synaptic plasticity. We find that the net
effect of adaptation is subtle and heterogeneous. Depending on
adaptation mechanism and test stimulus, adaptation can either increase
or decrease coding accuracy.
We discuss the neurophysiological and psychophysical implications of the
findings and relate it to published experimental data.
\end{abstract}
\newpage
\section*{Introduction}

Adaptation, a reduction in the firing activity of neurons to prolonged
stimulation, is a ubiquitous phenomenon that has been observed in many sensory
brain regions, such as auditory \citep{cynader1993,nelken2004}, and
somatosensory \citep{haidarliu2000} cortices, as well as in primary visual
cortex
(V1) \citep{freeman1982,ferster1997,dragoi2008}. In addition to response
reduction, adaptation has been observed to deform the tuning curves
\citep{lennie1999,dragoi2000,ghisovan2008a,ghisovan2008b}, which is thought to
lead to perceptual after-effects, such as the tilt after-effect
\citep{jin2005,schwartz2007,kohn2007,series2009,schwartz2009}. 

Adaptation is
hypothesized to be functionally important for coding sensory information
\citep{kohn2007,schwartz2007}. On the single neuron level, the reduction in
firing
rate should reduce coding accuracy, since it will typically become harder to
discriminate two similar stimuli when firing rates are lowered. However, on the
population level the accuracy is determined not only by the firing rates, but
also by the correlations between neurons. In primary visual cortex it has been
observed that adaptation reduces correlations, which can improve population
coding accuracy \citep{dragoi2008}. However, it is unknown how accuracy is
affected when both changes in firing rates and changes in the correlations are
taken into account.

The coding accuracy of a population response can be analyzed using Fisher
Information, which gives an upper bound to the accuracy that any decoder can
achieve. Importantly, the Fisher Information takes the effects of correlations
in neural variability (so called noise correlations) into account. Depending on
the exact structure of the correlations, the Fisher Information can either
increase or decrease with correlation strength
\citep{oram98,abbott99,sompolinsky02,series2004,averbeck2006}. Therefore a
precise determination of the correlation structure is required, which is
difficult to accomplish experimentally since vast amounts of data are required.
Moreover, experimental probe stimuli that measure coding before and after
adaptation can induce additional adaptation, possibly obscuring
the effect of the adapter.

Here we study this issue using a computational model of primary visual cortex.
We implement a variety of possible adaptation mechanisms and study coding
accuracy before and after adaptation. We find that the precise adaptation
mechanism and its parameters lead to diverse results: accuracy can decrease but
under certain conditions
also increase after adaptation. The changes in tuning curves resulting in
changes in coding are typically opposite from the changes in correlations.

\section*{Methods}

\subsection*{Network setup}

The model of V1, Fig.~\ref{fig1}A, was a rate-based version of the
ring model \citep{benyishai1995,somers1995,qian2003}. The model did
not
pretend to be a full model of the visual system, but included essential cortical
mechanisms that can produce adaptation. In the model, each of the $N=128$
neurons was labeled by its preferred orientation $\theta$, which ranged between
$-90$ and $90$ degrees. The instantaneous firing rate $\bar{R}_{\theta}$ (spikes
per second) was proportional to the rectified synaptic current.
\begin{eqnarray}
\bar{R}_{\theta} & =\kappa [I]_{+} +b,\label{R}
\end{eqnarray}
where $[x]_+ = \max(x,0)$. 
The constant $b$ is the background firing rate and $\kappa$ represents the gain
to produce the firing rate $\bar{R}_{\theta}$ for a given input $I$. Parameter
values can be found in Table 1.
We also used smoother rectification functions \citep{rossum2008}; however, this
did
not affect the results. The above firing rate model had the advantage of being
much more efficient than a spiking model, while Fisher information in spiking
models \citep{series2004} and rate models \citep{spiridon2001} have been found
to be
qualitatively similar. Moreover, implementation of realistic variability across
a wide range of firing rates is still a challenge in spiking models (for a
review see e.g. \citep{barbieri2008}).

For each neuron the synaptic current had a triple angular dependence and evolved
in time according to
\begin{eqnarray}
\tau\frac{\partial I(\theta,\phi,\psi,t)}{\partial t}=-
I(\theta,\phi,\psi,t)+I_{\mathrm{ff}}(\theta,\phi)+I_{\mathrm{exc}}(\theta,\phi,
\psi,t)-
I_{\mathrm{inh}}(\theta,\phi,\psi,t)-
I_{\mathrm{sfa}}(\theta,\phi,\psi,t)\label{intot}\nonumber \\
\end{eqnarray}
where $\psi$ is the adapter angle, $\phi$ is the test stimulus angle, and $\tau$
is the synaptic time constant \citep{dayanB}.
The current consisted of 1) feed-forward excitatory input $I_{\mathrm{ff}}$ from
the
LGN, 2) excitatory $I_{\mathrm{exc}}$ and inhibitory $I_{\mathrm{inh}}$ input
mediated
by lateral connections with a center-surround (Mexican-hat) profile subject to
short-term synaptic depression, and 3) an adaptation current $I_{\mathrm{sfa}}$
which
described spike frequency adaptation.

\subsubsection*{Network connections}
The input $I_{\mathrm{ff}}$ represented feed-forward excitatory input from LGN.
Except for Fig.~\ref{fig8} it was assumed not to be subject to adaptation
\citep{boudreau2005}. It
was modeled as a Gaussian profile with periodic boundary conditions:
\begin{eqnarray}
I_{\mathrm{ff}}(\theta,\phi) & = &
a_{\mathrm{ff}}\left[\exp\left(-\frac{(\theta-
\phi)^{2}}{2\sigma_{\mathrm{ff}}^{2}}\right)+\exp\left(-\frac{(\theta-
\phi+180)^{2}}{2\sigma_{\mathrm{ff}}^{2}}\right)+\exp\left(-\frac{(\theta-\phi-
180)^{2}}{2\sigma_{\mathrm{ff}}^{2}}\right)\right] \label{inff}\nonumber \\
\end{eqnarray}
where $\sigma_{\mathrm{ff}}$ represents the Gaussian width of stimulus
profile and $a_{\mathrm{ff}}$ its amplitude.

Lateral connections mediated the excitatory ($I_{\mathrm{exc}}$) and
inhibitory ($I_{\mathrm{inh}}$) inputs to neuron $i$ according to
\begin{eqnarray}
I_{\mathrm{exc}}(\theta_i,\phi,\psi,t) & = & g_{\mathrm{exc}} \sum_j
\mathcal{E}(\theta_i,
\theta_j)
x_{\mathrm{exc}}(\theta_j,\phi,\psi,t)R_{\theta_j}(\phi,\psi,t)\, \nonumber \\
I_{\mathrm{inh}}(\theta_i,\phi,\psi,t) & = & g_{\mathrm{inh}}\sum_j
\mathcal{I}(\theta_i,
\theta_j)x_{\mathrm{inh}}(\theta_j,\phi,\psi,t)
R_{\theta_j}(\phi,\psi,t)\,\label{inlat}\end{eqnarray}
where $\theta_j$ and $\theta_i$ represented respectively the pre- and
postsynaptic neurons, the constants $g$ were recurrent gain factors,
and $R_{\theta}(\phi,\psi,t)$ was a noisy realization of the firing
rate (see below). The variables $x$ in Eq.~(\ref{inlat}) represented
the efficacy of the synapses subject to short-term plasticity (see below as
well).

The functions $\mathcal{E}$ and $\mathcal{I}$ in Eq.~(\ref{inlat})
define the connection strength between cells $\theta_i$ and $\theta_j$.
We use \citep{qian2003}
\begin{equation}
\mathcal{E}(\theta_i, \theta_j) = [CK(|\theta-\theta'|)]_+
\label{E}
\end{equation}
and
\begin{equation}
\mathcal{I}(\theta_i, \theta_j) = [-CK(|\theta-\theta'|)]_+ 
\label{I}
\end{equation}
where\begin{eqnarray}
K(\theta) & = & \left[\cos\left(2\theta \right)+
1\right]^{A_{\mathrm{exc}}}-\left[\cos\left(2
\theta
\right)+1\right]^{A_{\mathrm{inh}}},\label{K}\end{eqnarray}
The functions $\mathcal{E}$ and $\mathcal{I}$
were normalized with the constant $C$ so that the sum of connections from any
cell to the rest
equals $1$, that is $1/C = \sum_i |K(\theta_i)|$.

The exponents $A_{\mathrm{exc}}$ and $A_{\mathrm{inh}}$ control
the range of interaction; the smaller they are, the flatter and wider
the connection profile.
We used a center-surround or Mexican-hat
profile for the connection, satisfying $A_{\mathrm{exc}}>A_{\mathrm{inh}}$,
as in previous studies \citep{benyishai1995,qian2003,series2004}.
The precise profile of lateral inhibition is not well known. However, we found
that 
in order to obtain sharpening (from a broad thalamocortical input to a sharp
cortical neuron response, Fig.~\ref{fig1}B) the shape of inhibition was
not critical,
i.e. both long range inhibition or a flat profile gave similar results.

\subsubsection*{Synaptic short-term plasticity}
In the model cortical synaptic short-term depression affected the lateral
connections only;
the afferents from the LGN did not depress \citep{boudreau2005}, Fig. S1C.
We used a phenomenological model of short-term depression \citep{tsodyks1997}:
The fraction of available neurotransmitter $x$ of a syanpse of a presynaptic
neuron
with orientation $\theta$ was
\[
\frac{\partial x(\theta,\phi,\psi,t)}{\partial t} = \frac{1-
x(\theta,\phi,\psi,t)}{\tau^{\mathrm{rec}}} -U 
x(\theta,\phi,\psi,t) R_{\theta}(\phi,\psi,t)\label{X}
\]
where $U$ is the release probability. The variable $x$ ranged between zero and
one, with zero corresponding to a fully depleted synapse and one to a
non-depressed synapse. Before stimulation and in the control condition, all $x$
were set to one. The constant $\tau^{\mathrm{rec}}$ was the recovery time from
depletion to resting conditions. During prolonged stimulation, the steady state
of the depression variable is smaller when the recovery time is slower, because
$x_{\infty}=\frac{1}{1+U\,\tau^{\mathrm{rec}} R}$.
We only included depression in the excitatory connections
\citep{galarreta1998,qian2003,chelaru2008a}. We found that the opposite
situation, namely stronger depression of inhibitory connections, lead to a
unstable network activity.

\subsubsection*{Spike frequency adaptation}
Spike frequency adaptation was implemented using a
standard first-order model \citep{herz2003}
\[
\tau_{\mathrm{sfa}}\frac{\partial I_{\mathrm{sfa}}
(\theta,\phi,\psi,t)}{\partial t} = -
I_{\mathrm{sfa}}(\theta,\phi,\psi,t)+g_{\mathrm{sfa}}R_{\theta}(\phi,\psi,
t).\label{sfa}
\]
The time constant $\tau_{\mathrm{sfa}}$ was set to 50ms and the gain
$g_{\mathrm{sfa}}$
was chosen so that the firing-rate reduction of the neuron to the adapter
stimulus is comparable to that produced by synaptic depression.

\subsection*{Variability model}

The responses of cortical neurons are typically highly variable, e.g.
\citep{tolhurst83}. To fit our
noise model we used extracellular single unit recordings from neurons in area V1
of macaque presented with 300 ms stimuli \citep{oram99}. For each recorded cell,
the
grating orientation that elicitepd the maximal response was labeled as the
"preferred orientation" (peak), and all orientations were given relative to the
peak (22.5, 45, 67.5 and 90 degrees). For n=19 different cells, 8 different
orientations were tested with 3 stimulus types: gratings, bars lighter than
background, bars darker than background. The spike rate was counted in 50 ms
bins. The firing rate for each orientation was calculated as the arithmetic mean
across the different stimulus types. In addition, symmetric orientations (e.g.
peak at +22.5 degrees and peak at -22.5 degrees) were collapsed, yielding 60
trials for the peak and the orthogonal orientation and 20 trials for all other
orientations. The variability was expressed in terms of the Fano Factor (FF),
which is the spike count variance across trials divided by the mean spike count.
We found that the FF was to a good approximation
constant across time and independent of stimulus orientation, 
nor did the Fano Factor depend on the firing rate, implying that the noise was
multiplicative.

To model this variability we added multiplicative noise to each neuron's
activity. The
noise is added after its input current is converted to the firing rate
Eq.\ref{R}, and noise reflects the stochastic nature of spikes generation. We
set
 \[
R_\theta=\bar{R}_\theta+\sigma(\bar{R}_\theta)\eta(t),\label{Reta}
\]
where the noise $\eta$ was Gaussian white noise, i.e. $\langle\eta(t)\rangle=0$,
and $\langle\eta(t)\eta(t')\rangle=\delta_{t,t'}$, and $\bar{R}_\theta$ denotes
the average
over all the trials. Note that the noise is temporally uncorrelated, reflecting
the approximate Poisson-like nature of cortical activity. Note that although
the added noise is
temporally and spatially uncorrelated, the synaptic coupling between the neurons
will
mix the noise from different neurons, resulting in correlated noise.

To achieve a given Fano Factor in the model,
the required amplitude of the noise was set according to
 $ \sigma(\bar{R}_\theta)=\sqrt{\mathrm{FF}}\sqrt{\bar{R}_\theta}.$
A sample trial is illustrated in Fig.~\ref{fig1}B (red dots).
To study how the trial-to-trial variability of the single cell affects
the population coding, we ran simulations using FF$=0.50,1.00,1.50$,
Fig.~\ref{fig5}C-D.
In addition, to further investigate the generality of our results we used two
alternative noise models: 
1) Poisson distributed firing rates with mean $\bar{R}$ (unity Fano Factor),
and 2) additive Gaussian white noise. We found that, although the absolute
scale of the Fisher Information differed, the adaptation affected the 
information exactly identically for all test angles.
In other words the effect of the adaptation was independent of exact noise
model.

For the common input noise model, Fig. \ref{fig7},
the feedforward input was corrupted with a multiplicative noise source
that was identical on a given trial for all neurons $I_{\mathrm{ff}}
(\theta,\phi)= (1+\eta) \bar{I}_{\mathrm{ff}} (\theta,\phi)$, 
where $\eta$ was a Gaussian random number with zero mean and unit variance.

\subsubsection*{Correlated variability}

In the visual cortex the response variability is correlated
\citep{nelson1992,gawne1996,kohn2005}. In the model these correlations result
from the lateral connectivity \citep{series2004}. 
The covariance matrix, Fig. \ref{fig3} top row, across trials was defined as
$$q_{ij}(\phi,\psi)\equiv\left\langle
\left[R_{i}(\phi,\psi)-\bar{R}_{i}(\phi,\psi)\right]\left[R_{j}(\phi,\psi)-
\bar{R}_{j}(\phi,\psi)\right]\right\rangle
_{\mathrm{trials}},$$
where $R_{i}=R_{\theta_{i}}$ was the firing rate of neuron $i$
and $\bar{R}$ the average across trials.
Although only the $q$-matrix enters
in the Fisher Information, experimental data is often reported using
normalizef correlation coefficients.
The (Pearson-r) correlation coefficient $c_{ij}$ was obtained by
dividing the covariance matrix
by the variances, $c_{ij} =
q_{ij}/[ \sqrt{\mathrm{var} (R_i) }\sqrt{\mathrm{var}(R_j)}]$, Fig. \ref{fig3}
bottom row.

The Mexican-hat connectivity produced a covariance matrix with both
positive and negative correlation, corresponding to excitatory (short range)
and inhibitory (long range) lateral connections,
similar to other models \citep{spiridon2001,series2004}.

\subsection*{Accuracy of population coding}

To quantify the accuracy of the population code we used the Fisher
Information, which is a function of the tuning
of individual neurons and the noise correlations, which for
Gaussian distributed responses equals
\[
\mathrm{FI}(\phi,\psi) = 
\mathrm{FI}_{1}(\phi,\psi)+\mathrm{FI}_{2}(\phi,\psi).\label{FI}\]
The first term is given by 
\begin{equation}
\mathrm{FI}_{1}(\phi,\psi) = 
\sum_{i}\sum_{j}\bar{R}'_{i}(\phi,\psi)q_{ij}^{-1}(\phi,\psi)\
\bar{R_{j}'}(\phi,\psi)\label{FI1}
\end{equation}
where $\bar{R}'$ denotes first derivative of $\bar{R}$ with respect to the
stimulus orientation $\phi$, and $q_{ij}^{-1}$ is the inverse covariance matrix.
Note that in contrast to population coding studies with homogeneous tuning
curves where the FI is constant, here
the FI depends both on the stimulus angle of adapter $\psi$ and the stimulus
angle of the probe.
The second term equals
\begin{equation} \mathrm{FI}_{2}(\phi,\psi) = 
\frac{1}{2}\sum_{i,j,k,l}q'_{ij}(\phi,\psi)q_{jk}^{-1}(\phi,\psi)q'_{kl}(\phi,
\psi)q_{li}^{-
1}(\phi,\psi)\label{FI2}
\end{equation}
For a covariance matrix independent of the stimulus orientation
$q'(\phi,\psi)=0$, so that 
$\mathrm{FI}_{2}(\phi,\psi)=0$.

To calculate the derivatives appearing in Eqs.(\ref{FI1}) and (\ref{FI2}),
we discretized the stimulus space with a resolution of $h=180/N$, with $N$ the
number of neurons.
The first derivative of the tuning curve of the neuron $i$ was computed as
$R'_{i}(\phi,\psi)=[R_{i}(\phi+h,\psi)-R_{i}(\phi-h,\psi)]/(2\,h)$,
and similar for the derivative of the correlation matrix in Eq.~(\ref{FI2}).
To accurately estimate the Fisher Information we used 12,000 trials for each
data point, as 
too few trials lead to an overestimate of the Fisher Information (by some 5\%
when using only
4000 trials).

Both contributions to the Fisher Information, given by Eqs.~\ref{FI1} and
\ref{FI2},
were affected by visual adaptation. Interestingly, both showed strikingly
similar
effects of adaptation, Fig.~\ref{fig5}A-B. Although quantitatively the $FI_2$
was
much smaller than $FI_1$ for the given number of neurons, the second term is
extensive
in the number of neurons, even in the presence of correlations, in contrast to
the
$FI_1$ term which saturates when correlations are present \citep{shamir2004}.
Because
the almost
identical effects of adaptation on the two terms, the effect of adaptation on
the 
sum of the two terms will be independent on their relative contribution, 
suggesting that our results will be valid for a wider range
of neuron numbers. 

The Fano Factor dependence for both terms can be analytically calculated;
applying the Fano Factor definition, the covariance matrix is
$q_{ij}(\phi,\psi)=\mathrm{FF}_{i}\bar{R}_{i}(\phi,\psi)\delta_{i,j}
+c_{ij}(\phi,\psi) \sqrt{\mathrm{FF}_{i}
}\sqrt{\mathrm{FF}_{j}}\sqrt{R
_{i}(\phi,\psi)}\sqrt{R_{j}(\phi,\psi)}$, where $c_{ij}$ are
pairwise correlations.
For a population of neurons with the identical FF (as is the case here),
$FI_{1}$ scales with $1/$FF, while $FI_{2}$ is independent of FF for our noise
model; this is illustrated
for the control condition in Fig.~\ref{fig5}, bottom row.

\subsubsection*{Simulation protocol}
The adaptation protocol is illustrated in Fig.~\ref{fig1}C. For each
trial, we ran the network without input until the neural activity stabilized
($\sim 150$ms),
after which an adapter stimulus with orientation $\psi$ was presented for 300
ms during which the network adapted. 
After this adaptation period, the adaptation state of the network was
frozen, i.e. no further adaptation took place. The response to a stimulus
with orientation $\phi$ was tested, which again required an
equilibration of the network activity (450ms). The activity at the end of
this period corresponded to a single trial response.

\subsection*{Decoders}

We measured how adaptation modified the performances of two decoders or
estimators:
a Population Vector decoder and a winner-take-all decoder
\citep{jin2005}.
For winner-take-all decoding the estimate was simply the (non-adapted) preferred
orientation 
of the neuron with highest firing rate. For the population vector decoder the
responses of all
neurons were vector summed with an orientation equal to the neuron's
preferred orientation before adaptation.

For each decoder, we computed both the bias $b(\theta)$ and the discrimination 
threshold $T(\theta)$. The bias is the difference
between the mean perceived orientation and the actual presented stimulus
orientation. 
The discrimination threshold 
follows from the trial-to-trial standard deviation in the
estimate, $\sigma(\theta)$ as \begin{equation} T(\theta)=\frac{D
\sigma(\theta)}{1+b'(\theta)} \end{equation} where $D$ is the discrimination
criterion,
chosen to be 1, corresponding to an error rate of $\sim 76\%$. The Cramer-Rao
bound on
the discrimination threshold is given by \citep{series2009} 
\begin{equation} T(\theta)
\geq \frac{D}{\sqrt{FI(\theta)}}.
\end{equation}

\section*{Results}

\begin{figure}[!ht]
\begin{center}
\includegraphics[width=12cm]{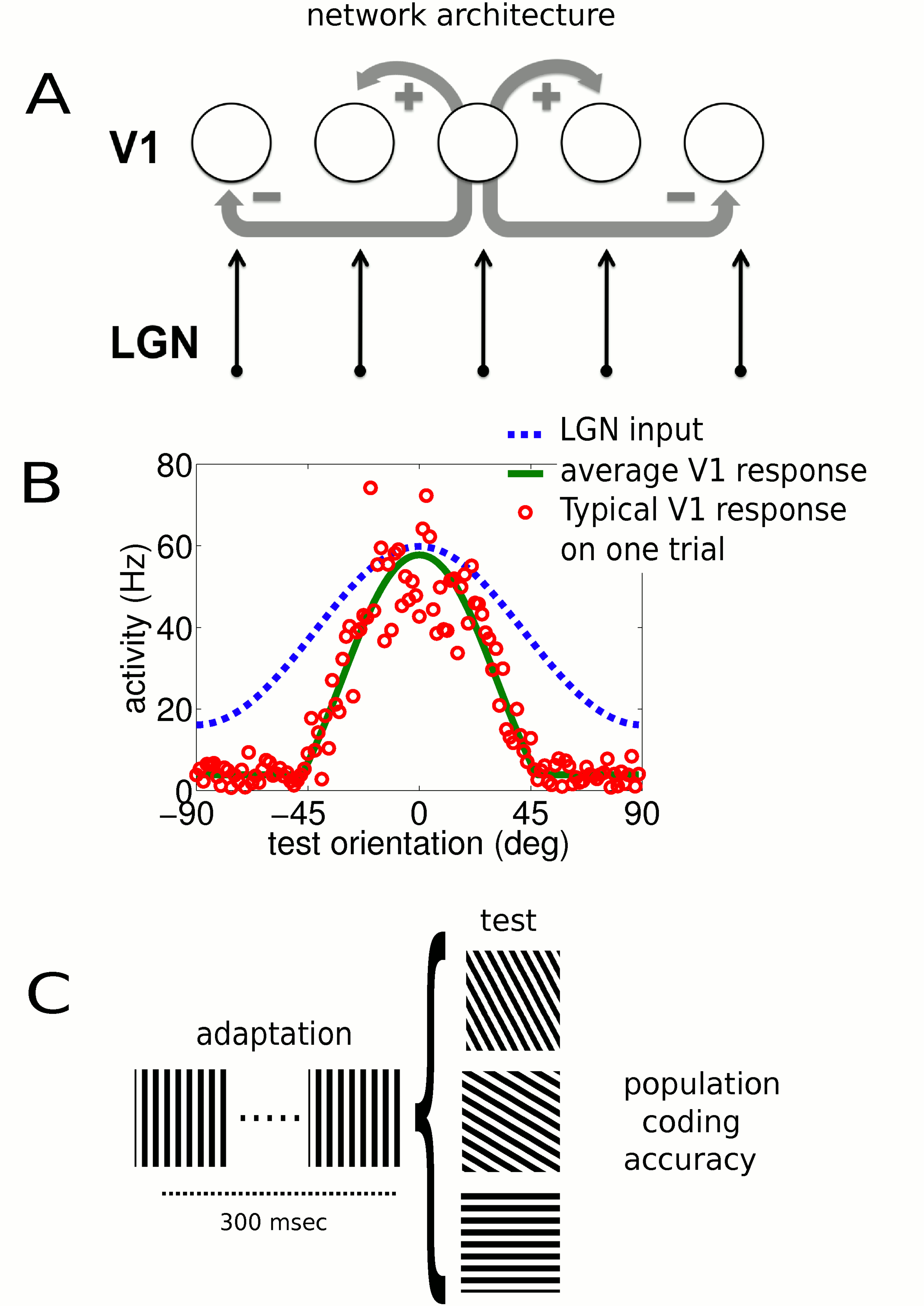}
\end{center}
\caption{
{\bf Network architecture, response and adaptation protocol.} 
A: The network architecture consists of a
recurrent network in which neurons
receive feedforward input from the LGN and lateral input with short range
excitation and long range inhibition
(Mexican hat profile). Periodic boundary conditions (not shown) ensured a ring
topology of the network.
B: Model behavior:
The cortical neurons (green line) sharpened the input from LGN (blue
line; scaled 15 times for clarity). 
On a single trial the multiplicative noise leads to substantial variability (red
dots).
 C: Schematic of the
adaptation protocol. During 300 ms the network is adapted to a stimulus
with orientation $\psi=0$, followed by a test phase with stimulus with 
orientation $\phi$ during which no further adaptation 
takes place.
}
\label{fig1}
\end{figure}

To study how adaptation affects coding, we used a well studied recurrent
network model of area V1, the so called ring model
\citep{benyishai1995,somers1995,qian2003}. Neurons in the network were
described 
by their firing rate and received feedforward input from the LGN and lateral
input via a center-surround profile, Fig.~\ref{fig1}A-B and Methods. 

Neural adaptation occurs through multiple mechanisms. Here we
focused on two distinct mechanisms both believed to contribute significantly to 
the adaptation of the neural responses to prolonged stimulation: 1) Spike
frequency adaptation (SFA), in which the neuron's firing rate decreases during
sustained stimulation \citep{mccormick2000a,mccormick2000b,herz2003}, and 2)
short-term
synaptic depression (SD) of the excitatory intracortical connections, so that
sustained
activation of the synapses reduces synaptic efficacy
\citep{finlayson1995,varela1997,nelson2002,best2004}. Synaptic
depression is comparatively strong in visual cortex as compared to
other areas \citep{wang2006}.

In the adaptation protocol, Fig.~\ref{fig1}C, an adaption inducing stimulus with
an angle $\psi$
was presented for 300 ms, after which the population response to a test stimulus
with orientation $\phi$ was measured. To insure that the models were
comparable, the parameters were set such that the adaptation mechanisms
reduced the firing rates of
neurons tuned to the adapter orientation equally, see Fig.~\ref{fig2}A-B.

\subsection*{Effect of adaptation on single neuron tuning }

First we studied how the different adaptation mechanisms changed the single
neuron tuning curves. Note that in homogeneous ring networks the response of a
neuron before adaptation is a function of a single variable only, namely the
difference between the stimulus angle $\phi$ and the neuron's preferred angle
$\theta$. Adaptation, however, renders the system inhomogeneous and the
responses were characterized by three angles: The angle of the adapter
stimulus $\psi$, set to 0 degrees, the angle of the test stimulus, and the
preferred orientation of the neuron. (Note that the
preferred orientation of a neuron refers to its preferred orientation before
adaptation).

\begin{figure}[!ht]
\begin{center}
\includegraphics[width=12cm]{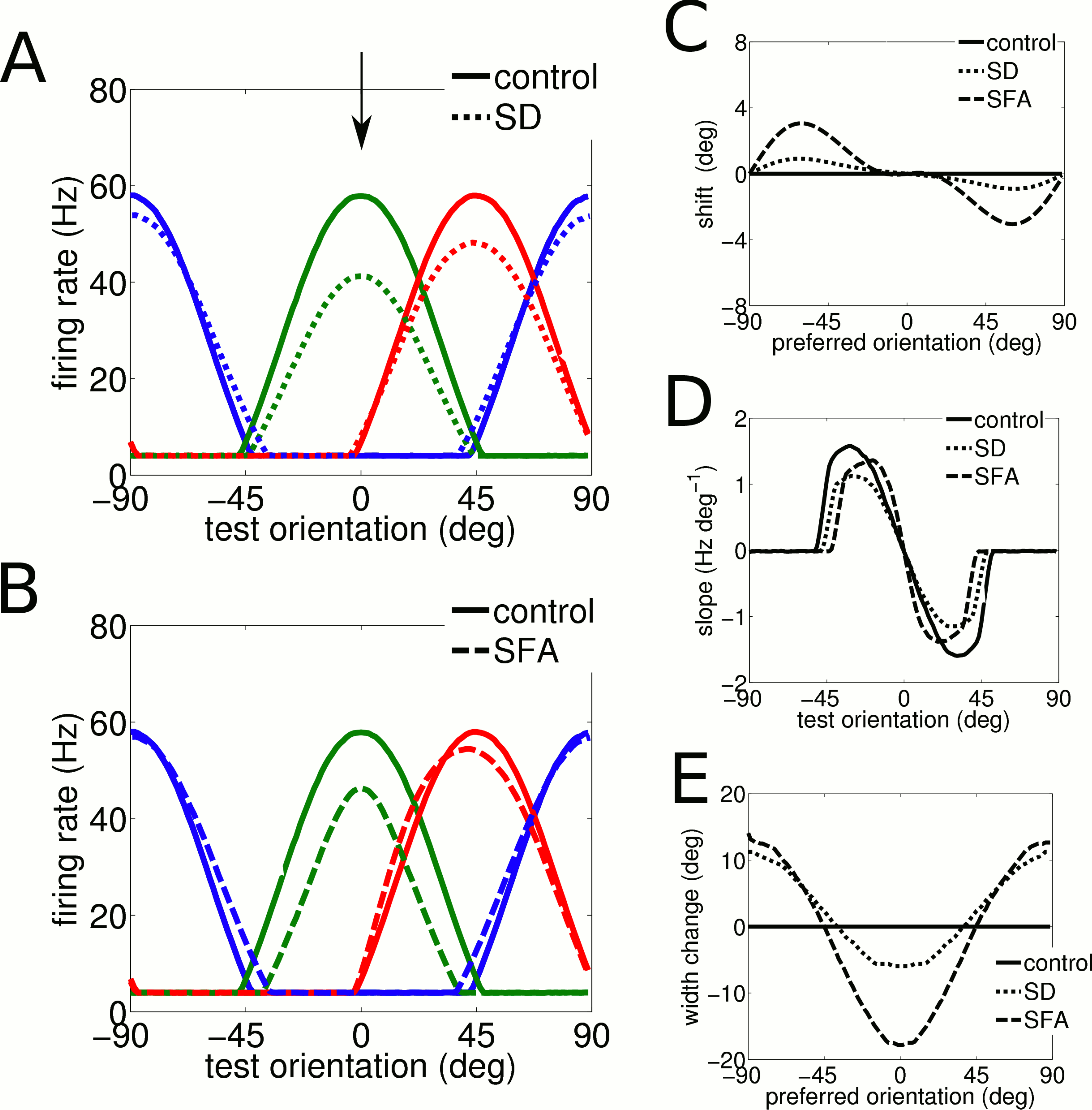}
\end{center}
\caption{{\bf Effect of adaptation on the tuning curves of individual neurons.} 
A-B Tuning curve properties for the different adaptation mechanisms:
synaptic depression (SD) and spike-frequency-adaptation (SFA).
 Individual neuron tuning curves before and after adaptation.
The tuning curves are shown for neurons with a preferred orientation of 0
(green), 45 (red) and 90 (blue) degrees, before (solid lines) and 
after adaptation (dashed lines).
The adapter orientation was set to zero degrees (marked with an arrow in panel
A).
C-E Changes in tuning curves properties after adaptation.
C: Postadaptation tuning curve shift (center of mass) for all neurons.
Both SD and SFA yielded an repulsive shift towards the adapter.
D: The slope of the tuning curve of a neuron with preferred orientation of zero
degrees
(green curve in panels A-B) as a function of the test orientation.
The absolute value of the slope at small test angles increased with
SFA, whilst the slope decreased with SD adaptation.
E: The change in the tuning curve width for the neurons in the population with 
respect to the control condition.
For both SD and SFA the tuning curves narrowed
for neurons with a preferred angle close to the adapter orientation,
and widened far from the adapter orientation.
}
\label{fig2}
\end{figure}

Both SD and SFA reduced the responses of neurons at the adapter
orientation, in addition the tuning curves deformed, see 
Fig.~\ref{fig2}A-B. We characterized the effect of adaptation on preferred
orientation shift, tuning curve slope and tuning curve width:

\textit{Tuning curve shift (Fig.~\ref{fig2}C)}: Adaptation shifts the
tuning
curve (characterized by the center of mass). For
both SD and SFA the tuning
curves shifted towards the adapter stimulus. Attractive shifts after adaptation
have
been reported before in V1\citep{ghisovan2008a,ghisovan2008b}, and in
area MT \citep{kohn2004}, although repulsive shift have been observed
in V1 as well \citep{lennie1999,dragoi2000}.

\textit{Tuning curve slope (Fig.~\ref{fig2}D):} The derivative of the
firing rate with respect to stimulus angle, or slope, is an important
determinant of
accuracy. In the absence of correlations the Fisher information is
proportional to the
slope (see Eq.~(\ref{FI1}), Methods). We plotted the slope of the tuning curve
as a
function of the test orientation for a neuron responding preferentially to the
adapter orientation. For test angles close to the adapter orientation ($<$20
degrees), SD reduced the absolute value of slope relative to the control
condition (solid line), but SFA increased the slope. Thus, for this neuron the
increase in the slope occurring for SFA increases the coding accuracy around
those
orientations. For test angles further from the adapter orientation (ranging from
45
to 57 degrees) both SD and SFA showed a decreasing slope.

\textit{Tuning curve width (Fig.~\ref{fig2}E):} The width was computed as the
angular range for which the neuron's response exceeded the background firing
rate.
(Similar results were obtained using the width at half height). For both SD and
SFA the width narrowed close to the adapter orientation
and widened at orthogonal angles. 

\subsection*{Effects of adaptation on noise correlations}

Coding accuracy is not only determined by the properties of the single neuron
tuning
curves, but also critically depends on the noise in the response and the
correlations
in the noise between the neurons \citep{abbott99,sompolinsky02,dragoi2008}.
These noise correlations describe the common fluctuations from trial to trial
between two neurons.
The noise was modelled by independent multiplicative noise to the firing
rates of the individual neurons,
mimicking the Poisson-like variability observed in vivo (Methods).
In the model direct and indirect synaptic connections lead to correlations in
the activity
and its variability. For instance, if a neuron is by chance more active on a
given trial, neurons which receive excitatory connections from it will tend to
be more active as well.

\begin{figure}[!ht]
\begin{center}
\includegraphics[width=14cm]{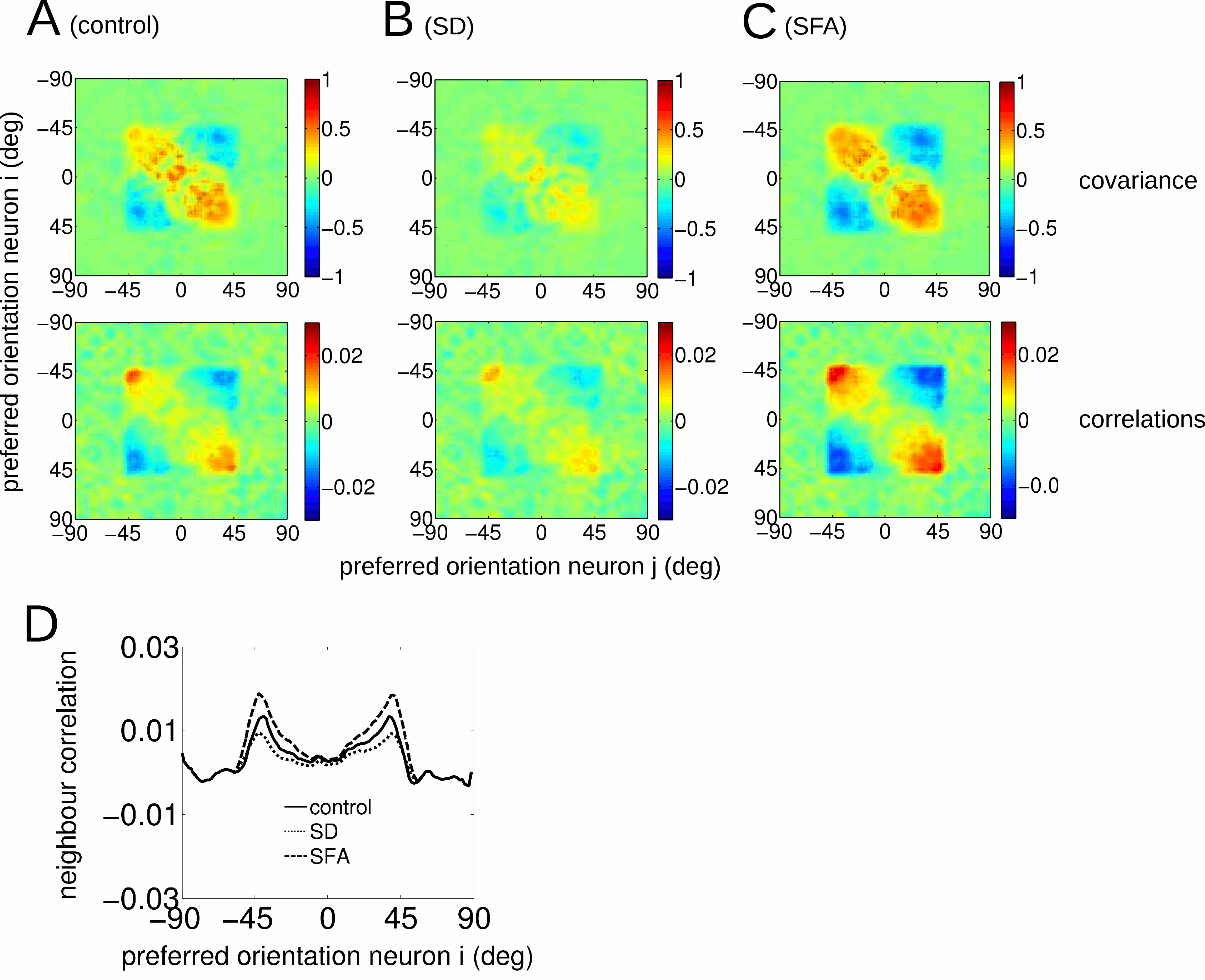}
\end{center}
\caption{
{\bf Effects of adaptation on the noise correlations.} On top row, the
covariance matrices are plotted 
such that 
each point at the surface represents the noise covariance between
two neurons with preferred orientation given by x and y-axis.
On bottom row, the covariance matrices were normalized to the noise (Pearson-r)
correlations coefficients.
The principal diagonal was omitted for clarity.
Compared to the control condition (panel A), synaptic depression (SD) 
reduced noise correlations (panel B), while
spike frequency adaptation (SFA) increased them (panel C).
D. The noise correlations coefficients depicted between neighboring cells (i.e.
$c_{i,i+1}$), for the three conditions. 
Adaptation through SD reduces the correlations, SFA increases them.
In all panels the adaptor and test orientation were set to 0 degrees.
}
\label{fig3}
\end{figure}

We first studied the correlations in the control condition. Fig.~\ref{fig3}A
top illustrates them for a stimulus angle of zero degrees. In analogy with
experiments we also characterized them by the Pearson-r correlation
coefficients. The correlation coefficients in the
control condition are shown in
Fig.~\ref{fig3}A, bottom. Only neurons with a preferred angle between
approximately -45 and +45 degrees were activated by this stimulus (the square
region in the center of the matrix), and could potentially show correlation. 

The correlations were strongest for neurons with similar preferred angles but
that responded quite weakly to the stimulus, while neurons that had
approximately orthogonal tuning (-45 and +45 degrees) showed negative
correlation, as observed previously \citep{spiridon2001,series2004}. The reason
is that the 
noise causes fluctuations in the position of the activity bump (the attractor
state) in the network, while its shape is approximately preserved.
In nearby neurons for which the stimulus was at the edge of
their receptive fields, such fluctuation lead to large, common
rate fluctuations and hence strong correlation. In contrast, the responses of
neurons in the center of the population, although noisy, were only weakly
correlated as the precise position of the attractor state hardly affected their
firing rates.

Adaptation led to changes in the noise correlation, Fig.~\ref{fig3}B-C,
bottom row. Spike frequency adaptation (SFA) increased noise correlations,
whilst synaptic depression (SD) decreased them. The latter is easy to
understand. Synaptic depression weakens the lateral connections, and thus
reduces correlations, shifting the operation of the network toward a feedforward
mode. 
Fig.~\ref{fig3}D shows the correlations between neighbouring neurons,
showing the shape of the correlations and the effect of the adaptation in one
graph.

\subsection*{Effect of adaptation on coding accuracy}

Next we combined the effects of adaptation on tuning curves and correlations to
calculate population coding accuracy as measured by the Fisher Information, see
Methods. The Fisher Information before and after adaptation is plotted as a
function of test orientation in Fig.~\ref{fig4}A. Before
adaptation the information is independent of the test angle as the network is
homogeneous (solid line). After adaptation, the information becomes dependent on
test angle (dashed curves). 

Only for SFA and only for test angles close to the adapter, the Fisher
Information and hence accuracy
increased. This peak was caused by the sharpening of the tuning curves near the
adapter angle (see below). For all other combination of test orientation and
adaptation mechanism the Fisher Information was slightly reduced. Interestingly,
the shapes of the curves for SD and SFA are similar in that the Fisher
Information peaked at the adapter angle. 

Next, we explored the effect of the correlations on the information. We shuffled
responses effectively removing any correlations. We subtracted the full Fisher
Information from the Fisher Information using shuffled responses,
Fig.~\ref{fig4}B. That this difference was always
positive indicates that correlations always reduced the accuracy. After SD
adaptation shuffling led to smaller increase than in control, consistent with
the
de-correlation caused by synaptic depression. For SFA the effect of
correlations was more dominant for orientations close to the adapter angle,
consistent with Fig.~\ref{fig3}C, and decreased for orientations far from the
adapter angle. 


As both individual tuning curves and the correlations change under adaptation,
we further distinguished their contributions to the Fisher Information. First,
we calculated the Fisher Information using adapted tuning curves and non-adapted
correlations, Fig.~\ref{fig4}C. Because the correlations were now fixed to the
control condition, changes in the
Fisher Information are a consequence of the changes in tuning curves only. The
Fisher Information resembled Fig.~\ref{fig4}A, indicating that
the deformations of the tuning curves are important. For SFA, the tuning curve
slope increased near to the adapter angle,
Fig.~\ref{fig2}D, resulting in a prominent peak in the Fisher
Information. 

Finally we calculated the Fisher Information using non-adapted tuning curves but
adapted correlations to reveal the information
changes due to changes in correlations only, Fig.~\ref{fig4}D. Consistent with
Fig.~\ref{fig3}B-C, with synaptic
depression the Fisher Information increased for most orientations as the
correlations are reduced.
Also for SFA, the Fisher Information increased at almost all stimulus angles,
except at a stimulus angle of zero degrees. In summary, when only
adaptation of the correlations was taken into account, the information increased
for most conditions. This finding
is consistent with recent experimental data \citep{dragoi2008}, where adaptation
of the noise correlations were shown to increase population accuracy.

Although we typically observed opposite effects of changes in tuning curves
and changes in correlation on the information (Fig.~\ref{fig4}C-D), the
total effect of the adaptation on coding is not simply
the linear sum of the two. They interact non-linearly and therefore the total
adaptation induced change of the Fisher Information in
Fig.~\ref{fig4}A is different from the sum of Fisher Information
changes from Fig.~\ref{fig4}C-D.

The Fisher Information can be split in two terms, $FI_1$ and $FI_2$
(see Methods). Both contributions to the Fisher Information, given by
Eqs.~\ref{FI1} and \ref{FI2}, were affected by visual adaptation. Interestingly,
both showed strikingly similar effects of adaptation, Fig.~\ref{fig5}A-B.
Although quantitatively the $FI_2$ was much smaller than $FI_1$ for the given
number of neurons, the second term is extensive in the number of neurons, even
in the presence of correlations. This is unlike the $FI_1$ term which saturates
when correlations are present \cite{shamir2004}. Because the almost identical
effects of adaptation on the two terms, the effect of adaptation on the sum of
the two terms will be independent on their relative contribution, suggesting
that our qualitative results will be valid for a wide range of neurons numbers. 

\subsection*{Reading out the population code}

Because it is not known how populations of neurons are read-out by the nervous
system, we analyzed how two well-known decoders, the winner-take-all and the
Population Vector decoder, are affected by adaptation (see Methods).
Importantly, if the decoder does not have access to the state of adaptation, 
as we assumed here, its
estimate will typically be biased \citep{series2009}. For the two decoders, the
bias profile was virtually identical, Fig.~\ref{fig6}A-B. Both SD and
SFA led to strong tilt after-effects (repulsive shift away from 
the adapter orientation). As observed before, the tuning of individual neurons
can behave 
differently from the population tuning \citep{qian2003}. The population bias can
be opposite from the individual tuning curve shift, cf.
Fig. \ref{fig6} to Fig. \ref{fig2}C. The reason is that the
population vector weighted individual responses according to their firing rate.
Therefore, although a neural tuning curve could shift towards the adapter
orientation, if the rate were sufficiently reduced, the population vector would
shift away from the adapter orientation.

For {\em unbiased} estimators the Cramer-Rao bound dictates that the minimum
obtainable variance is given by the Fisher Information. For biased estimators
where the bias is non-uniform this is no longer true. Instead, for biased
estimators, the Fisher Information forms a bound on the decoder's discrimination
threshold (see Methods and \cite{series2009}). Conveniently this is also the
quantity that is typically assessed psychophysically. The discrimination
threshold in the angle estimate in the two decoders is shown in
Fig.~\ref{fig6}C-D. The threshold is inversely proportional to the Fisher
information. Interestingly, for the winner-take-all decoder not only SFA but
also SD increase performance at zero degrees.
This is further illustrated in panel E-F, where the threshold is
plotted relative to the Fisher Information. This ratio is a measure of the
biased decoder efficiency, a value of one indicating a perfect decoder.
Unsurprisingly the population vector decoder (Fig.~\ref{fig6}E) is more
efficient than the winner-take-all (Fig.~\ref{fig6}F), which, unlike the
population vector, does not average over neurons to reduce fluctuations.

\subsection*{Effect of common input noise}

In our model the correlations were typically weak($<0.03$) and neurons with
orthogonal preferred angles were anti-correlated. With the exception of a recent
study \citep{ecker2010}, electrophysiological data in visual cortex, however,
show stronger correlations among nearby neurons with similar preferred
orientation ($\approx 0.2$) and typically less for neurons with different
preferred orientations \citep{nelson1992,gawne1996,deangelis1999,smith2008}. In
addition, pairwise correlations were found to be only weakly dependent on
stimulus orientation \citep{kohn2005}. This suggests that there is an additional
source of correlation in the data. In particular, common input noise from the
LGN input could be the source of this missing correlation. We implemented LGN
noise in the model that introduced common fluctuations in V1, enhancing the
noise correlation, Fig. \ref{fig7} right column. Although the common input noise
substantially increased the correlations, the effect on the Fisher Information
is negligible, cf. Fig.~\ref{fig7} bottom row. The limited effect of common
input noise on the coding accuracy can be understood from the fact that common
input noise can be extracted away by a decoder \citep{sompolinsky02}. A decoder
can first calculate the population averaged activity on each trial, and decode
relative to this average activity. 

\subsection*{Effect of input adaptation}

Although there is evidence that in the LGN input does not adapt
\citep{boudreau2005}, we also tested the case where the input did adapt through
synaptic depression. Parameters were adjusted so that the same response
reduction after adaptation was achieved as SD and SFA. The adaptation reduces
the drive to the V1 network. The adaptation effects on tuning curves and Fisher
Information are shown in Fig. \ref{fig8}. These results are very similar to SD.
However, in contrast to SD, the correlations before and after adaptation were
virtually identical (not shown), as adapting the input did not affect the
lateral connections. 

\section*{Discussion}

We built a recurrent network model of primary visual cortex to address how
adaptation affects population coding accuracy. Although a variety of models have
modeled
adaptation by imposing changes in the tuning curves, e.g. \citep{jin2005}, our
model is
the first to model the adaptation biophysically, thus directly relating
biophysical adaptation
to coding. From the adaptation mechanism, it predicts the
concurrent changes in tuning curves, correlations, and coding.
For synaptic depression (SD), we observed at the adapter orientation a strong
reduction of tuning curve width, attractive shifts of the individual tuning
curves, and strong repulsive population shifts. The case of spike frequency
adaptation (SFA) was qualitatively comparable to SD with regards to
tuning curves, but the changes in the coding accuracy were distinct (see
below).

Experimental data on tuning curves changes after adaptation are diverse and
sometimes contradictory. Increased widths at the adapter orientation and
repulsive
shifts have been reported in V1 \citep{dragoi2000}, while other studies in V1
reported a width reduction to
the adapter stimulus and repulsive shifts \citep{lennie1999}. Yet another study
reported attractive and
repulsive shifts, with no explicit
quantification of width variations of the tuning curves 
\citep{ghisovan2008a,ghisovan2008b}. Finally, electrophysiological
data from direction selective cells in area MT show a strong narrowing at the
adapter orientation and attractive shifts \citep{kohn2004}.

Next, we examined the effect of adaptation on coding accuracy by means of the
Fisher
Information. We studied how Fisher Information depends on test orientation and
precise adapting mechanism. The Fisher Information represented in
Fig.~\ref{fig4}A
showed larger Fisher Information for orientations close to the adapter
orientation in the case of SFA. This was due to the increased slope of the
tuning curves after adaptation. The increase in information occured despite
increased noise correlations.
In the case of SD adaptation the accuracy reduced, despite a reduction in the
correlations, Fig.~\ref{fig3}. This reduction was however not enough to counter
the reduction in accuracy due to tuning curve changes, Fig.~\ref{fig4}C.

We also considered the case where the input connections are
adapting. It should be noted that whether the LGN-V1 connections are subject to
short term synaptic depression is matter of debate
\citep{boudreau2005,Jia2004}.
Inclusion of synaptic depression in the LGN-V1 connections reduced the drive of
V1 for neurons close to the adapter.
The results for both the
tuning curve and Fisher Information changes is very comparable to SD, Fig.
\ref{fig8}.

To compare these findings to psychophysics we implemented two common decoders
which yield adaptation-induced biases and accuracy. Both decoders showed
repulsive tilt after-effects.
Both attractive and repulsive have been reported experimentally, typically
showing
repulsion close to the adapter and attraction far away
\citep{schwartz2007,clifford2000,schwartz2009}. Although we do find repulsion
close
to the adapter, attraction for orthogonal orientation is not reproduced here.
This
could be due either to differences in the activity or to a different readout
mechanism.
Data on orientation discrimination performance
after adaptation is limited, although it has been argued to resemble motion
adaptation
\citep{clifford2002b}, for which more data is available, see e.g.
\citep{series2009}. 
Similar to the case of SFA, two studies found a slight improvement around 0
degrees, but
a higher (worse) threshold for intermediate angles
\citep{regan1985,clifford2001}. One
study found a lower (better) threshold for angles close to 90 degrees
\citep{clifford2001}, not present in our simulations.

\subsection*{Model assumptions and extensions}

As any model, this model has certain limitations and is based on a number of
assumptions. First, the model is not spiking, but rate based. Of course a
spiking model is in principle more realistic and could lead to different
results, but apart from the difficulty tuning a spiking network model, in
particular obtaining the right noise and correlations is far from trivial, such
a model would be computationally challenging given the number of trials required
to calculate the Fisher Information. Encouragingly however, in studies of
non-adapting networks, a rate model with phenomenological
variability lead to a similar covariances as a network with spiking
neurons \citep{spiridon2001,series2004}. 

We assumed that the noise stems either from the neurons in the
network or from common LGN modulation. Additional sources of noise do
probably exists, including correlated noise sources such as other common input,
common feedback, or common modulation, both stationary and adapting. Although
such noises could be included in the model but too little is known about noise
sources to allow for strong conclusions on this point. 

Our model focusses on brief adaptation paradigms of up to a few hundred ms, as
cellular and synaptic adaptation is best characterized on those time scales. We
have modelled simplified versions of synaptic and cellular
adaptation that, although these phenomological models are well supported by
empirical data, are known to be much more complicated biophysically and
manifest multiple non-linearities and timescales. For
longer adaptation periods, other adaptation processes might be activated, e.g.
\citep{mccormick2000b,wark2007}, which would require modified adaptation
models. 
Despite these limitations, one could in principle consider cases where the
network is continuously stimulated, as would occur under natural stimulation.
Our results suggest that the coding accuracy would show a strong history
depedence. Furthermore, the net effect of the synaptic depression is expected to
be less, as the synapses will be in a somewhat depressed state already.
Unfortunately, a detailed analysis of this scenario seems currently
computationally prohibitive.

\begin{figure}[!ht]
\begin{center}
\includegraphics[width=12cm]{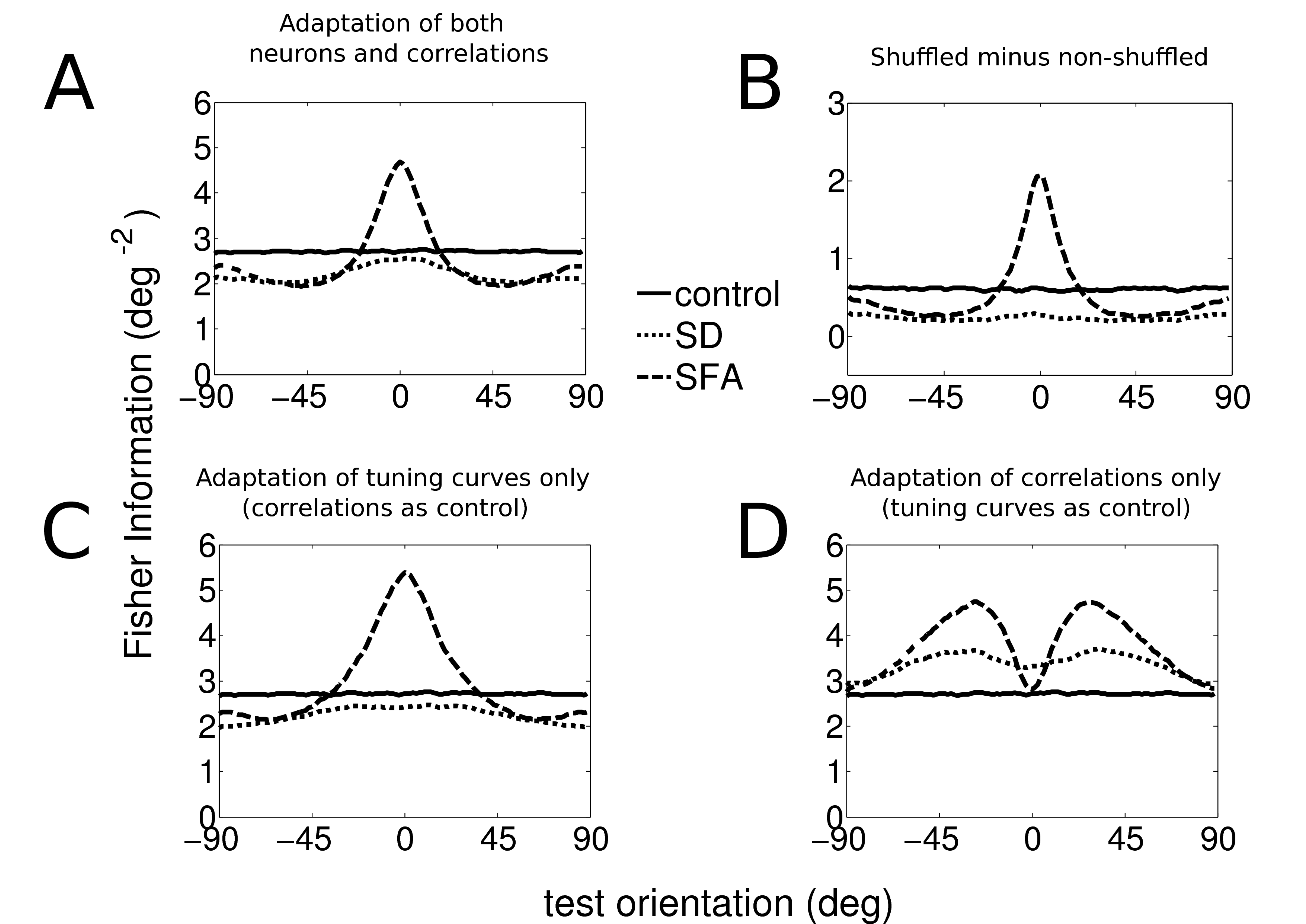}
\end{center}
\caption{
{\bf Effect of adaptation on population coding accuracy.}
 A: Fisher Information as a function of test orientation, in the control
condition, i.e. before
adaptation, and after adaptation through synaptic depression (SD) and spike
frequency adaptation (SFA).
Compared to the control condition the Fisher Information increases for SFA,
provided the test stimulus orientation was similar to the adapter orientation. 
In all others cases a slight reduction of the accuracy is seen.
B: The additional information when the noise correlation is removed by 
shuffling the responses. The difference of Fisher Information using
original responses (as panel A) and shuffled responses is shown. In all
cases, including control, shuffling increases the information.
For SD as after adaptation the population response is
partially decorrelated, 
less information is gained by shuffling than in the control condition.
For SFA the effect is heterogeneous,
indicating that adaptation effects of the noise correlations for SFA were high
for
orientations close to the adapter orientation and less dominant far from it.
C: Fisher Information using adapted tuning curves and non-adapted correlations.
D: Adapted correlations with non-adapted tuning curves. 
Using non-adapted tuning curves, the decorrelation for 
SD adaptation increased Fisher Information.
}
\label{fig4}
\end{figure}

\begin{figure}[!ht]
\begin{center}
\includegraphics[width=12cm]{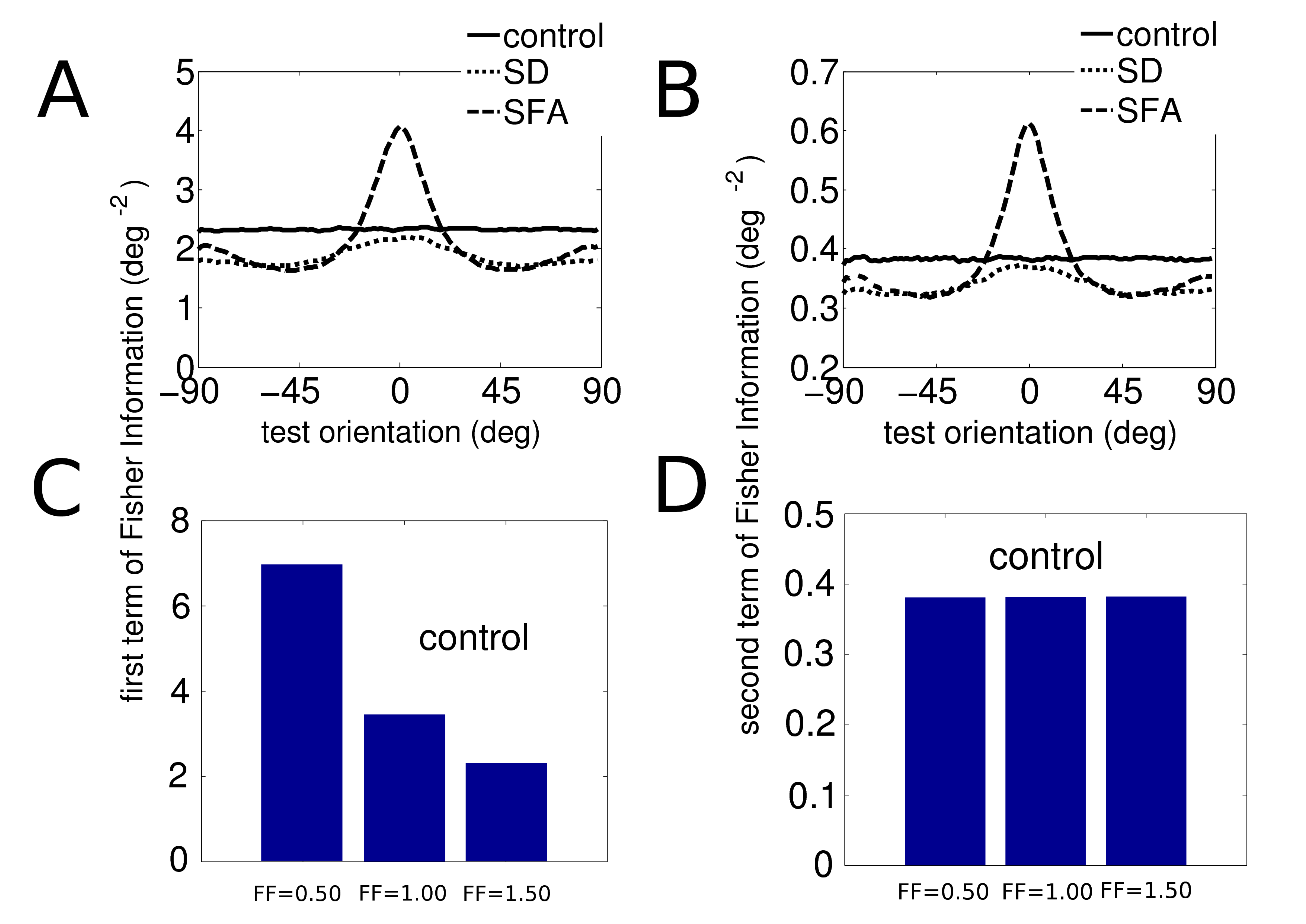}
\end{center}
\caption{
{\bf Contributions to the constituents of the Fisher Information.}
The Fisher Information is the sum of the term $FI_1$ (left) and $FI_2$ (right),
see Eqs.~\ref{FI1} and \ref{FI2}. A-B:
$FI_1$ and $FI_2$ for the adaptation scenarios
compared to the control situation. 
The effect of the adaptation dependence was similar, although the second term
was an order of magnitude 
smaller.
 C-D: Dependence of $FI_1$ and $FI_2$ on the Fano Factor (FF), using
 Gaussian multiplicative noise.
Only the control situation (before adaptation) is plotted, as the scaling before
and after
adaptation is identical.}
\label{fig5}
\end{figure}

\begin{figure}[!ht]
\begin{center}
\includegraphics[width=12cm]{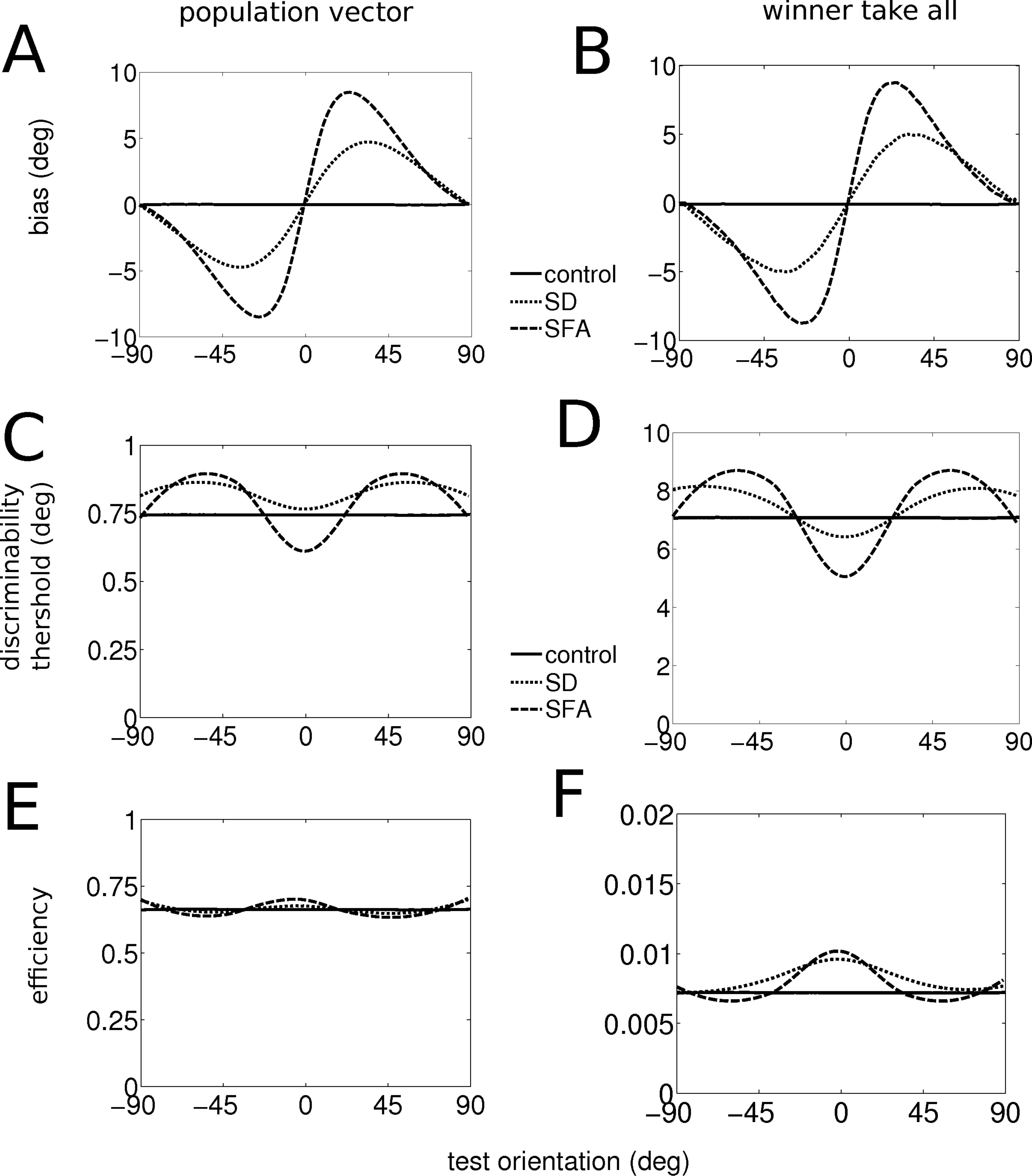}
\end{center}
\caption{
{\bf Decoding bias and variance.}
A-B: The bias (average estimated orientation minus actual orientation) for two
decoders: the population vector
 and winner-take-all decoder. SD and SFA adaptation induced a positive bias for
estimations of
positive orientations and negative bias for negative estimations
(i.e. a repulsive shift of the population). 
C-D: The discrimination threshold of the decoders is inversely proportional to
the inverse of the Fisher
Information, Fig.~\ref{fig4}A.
E-F: The discrimination threshold relative to the Fisher Information.
}
\label{fig6}
\end{figure}

\begin{figure}[!ht]
\begin{center}
\includegraphics[width=12cm]{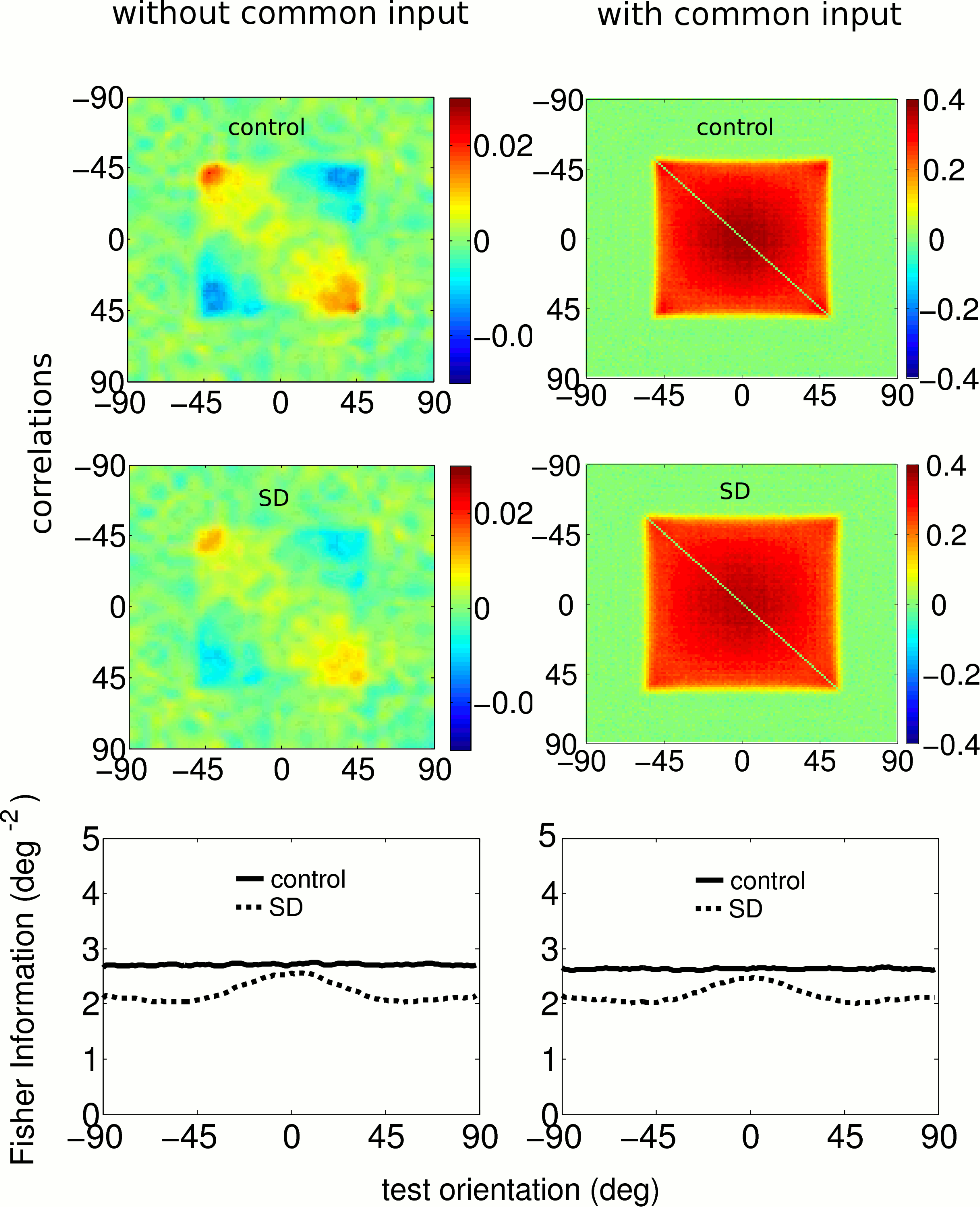}
\end{center}
\caption{
{\bf Effect of common input noise on correlation and coding.}
Compared to the situation without common noise (first column), after
the inclusion of common noise input (second column)
 the correlation structures changed significantly. 
With common input the correlations increased from less than 0.1 to 0.4, and the
correlations became
weakly dependent on the stimulus orientation.
Because the LGN input did not depress,
changes in the correlations were practically unaffected by adaptation. 
However, the common noise did not modify 
the information (bottom panels). For clarity, only adaptation trough SD is
shown; the same holds for SFA.
}
\label{fig7}
\end{figure}

\begin{figure}[!ht]
\begin{center}
\includegraphics[width=12cm]{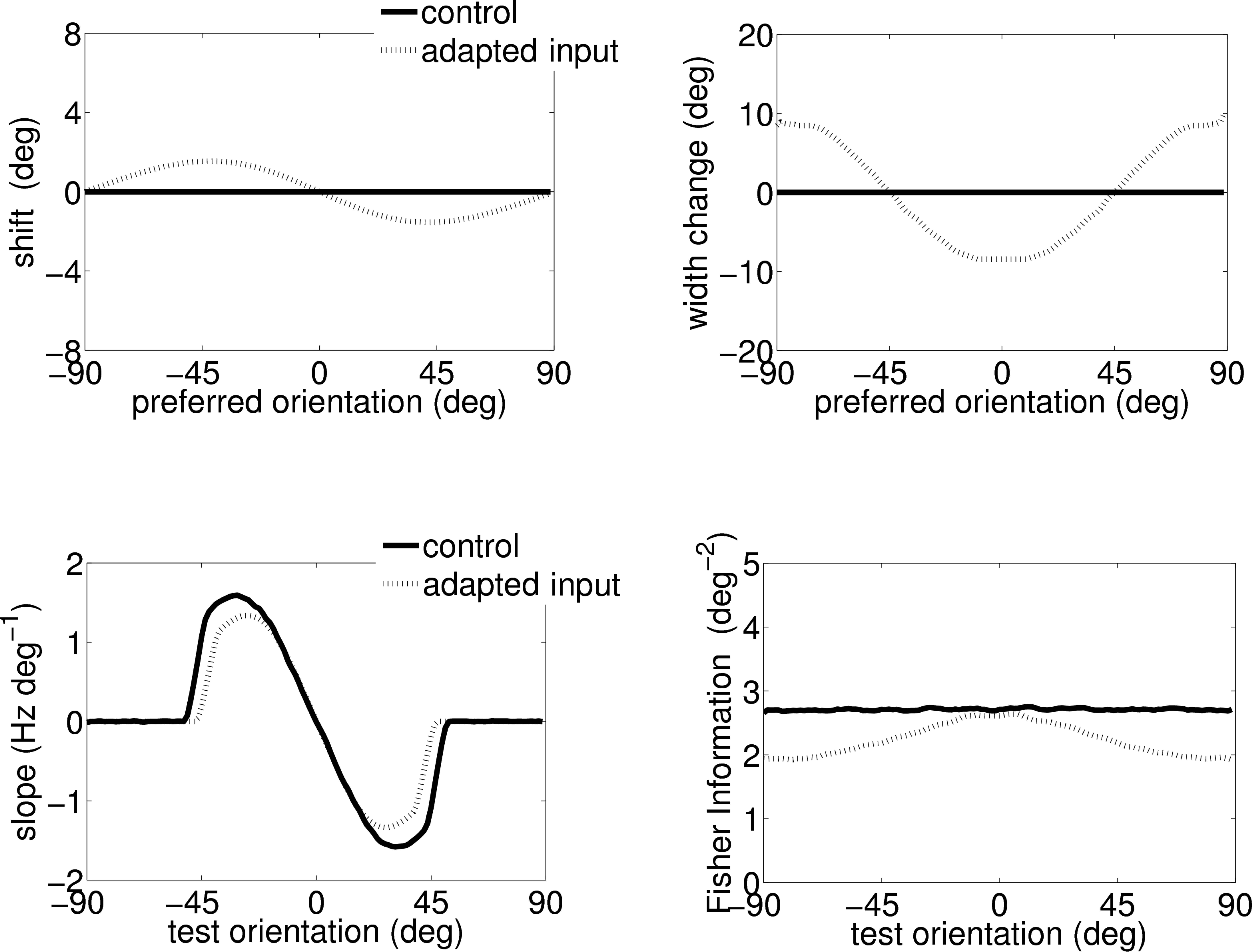}
\end{center}
\caption{ {\bf The effect of input adaptation on tuning curve properties and Fisher
Information}. In this simulation the input current
was adapted. The strength of adaptation was adjusted to have a similar
amount of reduction in the peak response. No other adaptation mechanism
was active. The changes in tuning curve and Fisher Information are
very similar to the changes seen with SD.
 }
\label{fig8}
\end{figure}

Finally, we assumed a model in which the tuning curves are sharpened by the
recurrent connections. Previous studies, without adaptation, have contrasted
coding in such networks with coding in feedforward networks with no or much less
recurrence and found that the recurrence induces correlations that strongly
degrade the coding \citep{spiridon2001,series2004}. Inline with those findings,
we find that the correlations induced by the recurrent connections, severly
reduce the information, see Fig. \ref{fig4}C. Whether the primary visual cortex
is
recurrently sharpened or dominated by feed-forward drive is a matter of debate
\citep{Ferster00}. In the feedforward variant of our model without any
recurrence, noise correlations will be absent. Thus
the tuning curves and the noise determine the information content and the
changes caused by adaptation. 
This situation is comparable to calculating the information using uncorrelated
(i.e. shuffled) responses, Fig. \ref{fig4}B.

In summary, adaptation strongly reduces neuronal activity but the post
adaptation changes in tuning curves and correlations and hence coding
accuracy depend strongly on the biophysical details of the adaptation
mechanism.

We thank to Adam Kohn, Odelia Schwartz, Walter Senn, Andrew Teich,
Andrea Benucci, Klaus Obermayer and Lyle Graham for useful comments and
suggestions.
This research was funded by EPSRC COLAMN project (JMC and MvR).
JMC was funded by the Fulbright
Commission to visit the CNL Lab at the Salk Institute. TJS is
supported by Howard Hughes Medical Institute. DM received
funding to visit Edinburgh from the High-Power-Computing
European Network (HPC-EUROPA Ref. RII3-CT-2003-506079). JMC is currently funded
by the Spanish Ministerio de Ciencia e Innovacion, through a Ramon y
Cajal Research Fellowship.
JMC thanks the computer facilities from the COLAMN
Computer Cluster (University of Plymouth), the Edinburgh Parallel
Computing Center and the Edinburgh Compute and Data Facility.


\clearpage
\newpage

\clearpage \newpage

\clearpage \newpage

\clearpage \newpage

\clearpage \newpage

\clearpage \newpage

\clearpage \newpage

\clearpage \newpage

\clearpage
\newpage

\section*{Tables}
\begin{table}[!ht]
\caption{{\bf Symbols and parameters used in simulations.} 
The parameters for SD and SFA were chosen to achieve the same control
conditions, and 
a similar response reduction after adaptation for the neuron maximally responding at the adapter
orientation. 
Finally all parameters were restricted to be in the physiologically plausible range.}
\begin{tabular}{|l|l|l|}
\hline
\textbf{Meaning} & \textbf{Symbol} & \textbf{Value} \tabularnewline
\hline
neuron's preferred orientation & $\theta$ & $(-90,90)$ degrees \tabularnewline
\hline
adapter orientation & $\psi$ & 0 degrees \tabularnewline
\hline
test orientation & $\phi$ & $(-90,90)$ degrees \tabularnewline
\hline
time decay of synaptic current & $\tau$ & 10 ms \tabularnewline
\hline
amplitude of thalamocortical input & $a_{\mathrm{ff}}$ & 4.0 \tabularnewline
\hline
Gaussian width of thalamocortical input & $\sigma_{\mathrm{ff}}$ & 45.0 \tabularnewline
\hline
gain for excitatory intracortical input & $g_{\mathrm{exc}}$ & 0.2 \tabularnewline
\hline
interaction power for excitatory connections & $A_{\mathrm{exc}}$ & 2.2 \tabularnewline
\hline
gain for inhibitory intracortical input & $g_{\mathrm{inh}}$ & 2.5 \tabularnewline
\hline
interaction power for inhibitory connections & $A_{\mathrm{inh}}$ & 1.4 \tabularnewline
\hline
probability of excit. transmitter release & $U$ & 0.02
\tabularnewline
\hline
recovering time for synaptic depression & $\tau$ & 600 ms \tabularnewline
\hline
background firing rate & $b$ & 4Hz \tabularnewline
\hline
input current to firing rate gain & $\kappa$ & 4.0 Hz \tabularnewline
\hline
Fano Factor & FF & 1.5 \tabularnewline
\hline
time decay for spike frequency adaptation (SFA) & $\tau_{\mathrm{sfa}}$ & 50 ms
\tabularnewline
\hline
gain for spike frequency adaptation & $g_{\mathrm{sfa}}$ & 0.05 \tabularnewline
\hline
\end{tabular}
\label{tableparameters}
\end{table}

\end{document}